\documentclass[%
reprint,
superscriptaddress,
 amsmath,amssymb,
 aps,
]{revtex4-2}

\usepackage{graphicx}
\usepackage{dcolumn}
\usepackage{bm}
\usepackage[hidelinks]{hyperref}
\usepackage{color}
\usepackage{siunitx}
\usepackage{physics}

\begin{document}

\preprint{APS/123-QED}

\title{
Sub-Recoil Transverse Momentum Width in a Cold Ytterbium Atomic Beam
}

\author{Toshiyuki Hosoya}

\affiliation{Product Development Center, Japan Aviation Electronics Industry, Ltd.,\\
3-1-1, Musashino, Akishima-shi, Tokyo, 196-8555, Japan}
\affiliation{Department of Physics, Institute of Science Tokyo,\\
2-12-10 Ookayama, Meguro-ku, Tokyo, 152-8550, Japan}

\author{Tomoya Sato}
\affiliation{Institute of Integrated Research, Institute of Science Tokyo,\\
4259 Nagatsuta-cho, Midori-ku, Yokohama, Kanagawa, 226-8503, Japan}

\author{Ryotaro Inoue}
\altaffiliation[Current affiliation: ]{Nanofiber Quantum Technologies, Inc.}
\affiliation{Institute of Integrated Research, Institute of Science Tokyo,\\
4259 Nagatsuta-cho, Midori-ku, Yokohama, Kanagawa, 226-8503, Japan}

\author{Mikio Kozuma}
\email{kozuma@qnav.iir.isct.ac.jp}
\affiliation{Department of Physics, Institute of Science Tokyo,\\
2-12-10 Ookayama, Meguro-ku, Tokyo, 152-8550, Japan}
\affiliation{Institute of Integrated Research, Institute of Science Tokyo,\\
4259 Nagatsuta-cho, Midori-ku, Yokohama, Kanagawa, 226-8503, Japan}

\date{\today}

\begin{abstract}
We demonstrate the generation of a slow ytterbium atomic beam with a transverse momentum width of $0.44(6)$ times the photon recoil associated with Bragg diffraction, and a flux of $6.7(9) \times 10^6$ atoms/s. This is achieved by applying momentum filtering through a long-lived metastable state to atoms prepared in a slow beam via two-dimensional transverse laser cooling. The resulting narrow momentum distribution enables efficient quasi-Bragg diffraction, which we exploit to realize a Bragg interferometer. These results mark a significant step toward continuous, high-precision, and magnetically insensitive angular rate measurements using cold alkaline-earth(-like) atomic beams.
\end{abstract}

\maketitle


Since the Sagnac phase shift induced by rotation is proportional to the energy of the particles involved~\cite{clauser_ultra-high_1988}, extensive research has focused on developing highly sensitive gyroscopes based on atomic interferometers as alternatives to conventional light-based systems. Atom-interferometry-based gyroscopes~\cite{riehle_optical_1991,gustavson_precision_1997,gustavson_rotation_2000} have a wide range of potential applications, including precise inertial navigation~\cite{titterton_strapdown_2004} and tests of relativistic theories such as the geodetic and Lense–Thirring effects~\cite{dimopoulos_general_2008,stockton_absolute_2011,he_space_2023}.
Particularly for inertial navigation, dead-time-free measurement, recently demonstrated using an atomic fountain interferometer~\cite{savoie_interleaved_2018}, is indispensable, as the vehicle's orientation must be continuously tracked. Cold atomic beam interferometers have been intensively investigated, as they enable not only continuous measurements but also high measurement bandwidth. Such capabilities are essential for realizing practical inertial navigation systems.
Recently, techniques for generating cold atomic beams using alkali atoms have been established~\cite{feng_observation_2015,kwolek_three-dimensional_2020}, enabling the development of atomic interferometers based on these beams~\cite{kwolek_continuous_2022,meng_closed-loop_2024}.
One of the next major challenges is to improve robustness against environmental magnetic fields to ensure high stability in field applications.
A promising approach is to generate cold atomic beams of alkaline-earth(-like) atoms such as strontium  or ytterbium (Yb)~\cite{nosske_two-dimensional_2017,wodey_robust_2021}, and to construct atomic interferometers using Bragg diffraction, taking advantage of the magnetic insensitivity of the $^1{\rm{S}}_0$ ground state.

The interaction time required for Bragg diffraction can be significantly reduced by employing quasi-Bragg diffraction with a Gaussian pulse~\cite{muller_atom-wave_2008}. This approach increases the momentum width over which diffraction can occur; however, achieving efficient Bragg diffraction still requires a momentum width narrower than the photon recoil~\cite{szigeti_why_2012}. Methods such as evaporative cooling~\cite{torii_mach-zehnder_2000} or momentum selection using quasi-Bragg diffraction itself, applied after Doppler cooling with ultra-narrow transitions~\cite{muller_atom_2008,mazzoni_large-momentum-transfer_2015}, have been employed to meet this requirement. Nonetheless, generating a continuous cold atomic beam using these methods remains challenging due to the long cooling times involved.
Mechanical slits can in principle yield a cold atomic beam with a narrow transverse momentum width~\cite{giltner_atom_1995}, but this method significantly restricts the atomic flux. Although state-of-the-art laser cooling with optical dipole traps can produce atoms with sub-recoil transverse momentum widths, the achievable flux remains limited to the order of $10^4$ atoms/s~\cite{chen_continuous_2022}. Alternatively, coherent population trapping techniques~\cite{aspect_laser_1988} can cool atoms without significant flux loss; however, they require sublevels in the ground state, which are absent in atoms with a zero magnetic moment, such as those considered here.

In this letter, we report the first realization of a slow Yb atomic beam with a sub-recoil transverse momentum width, achieved by momentum filtering through a long-lived metastable state applied to a cold beam prepared by two-dimensional transverse laser cooling~\cite{hosoya_high-flux_2023}. We also demonstrate higher-order quasi-Bragg diffraction and a Bragg interferometer utilizing this continuous atomic beam.

\begin{figure}[h]
\includegraphics[scale=0.42]{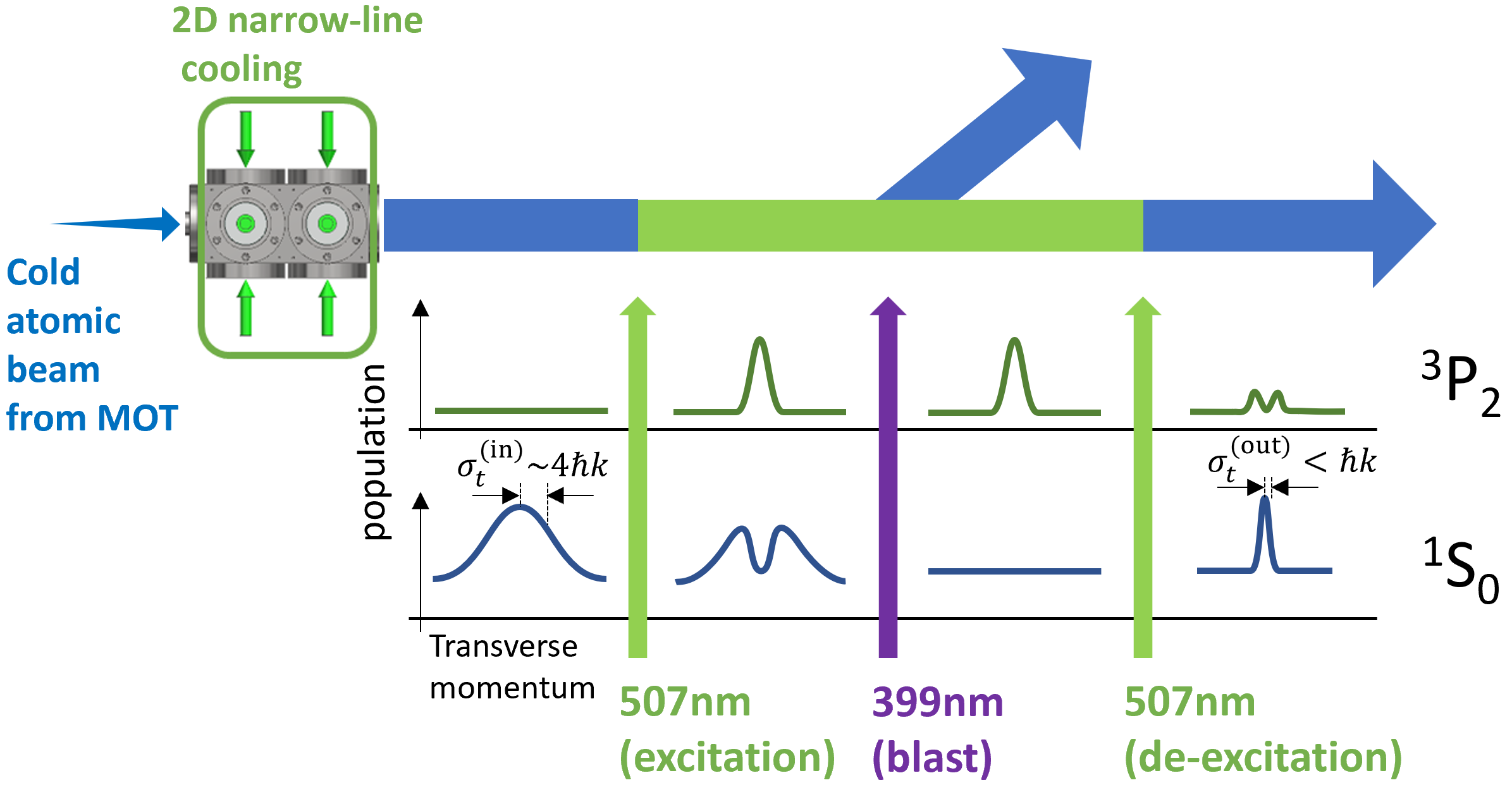}
\caption{\label{fig:fig1} The concept of a momentum filtering scheme.}
\end{figure}

Figure \ref{fig:fig1} describes the concept of a momentum filtering scheme. 
Although Yb is used for illustration, the method can be readily applied to other atomic species with suitable metastable states. An excitation laser beam resonant with the ultra-narrow
${}^{1}\mathrm{S}_0 - {}^{3}\mathrm{P}_2$ transition (wavelength:~\SI{507}{nm}, natural linewidth:~\SI{25}{mHz}~\cite{porsev_hyperfine_2004}) 
is introduced orthogonally to the atomic beam which has a transverse momentum width broader than the photon recoil $\hbar k_{399}$ of the Bragg light tuned to the dipole-allowed
${}^{1}\mathrm{S}_0 - {}^{1}\mathrm{P}_1$ transition (wavelength:~\SI{399}{nm}, natural linewidth:~\SI{29}{MHz}~\cite{takasu_photoassociation_2004}).
Here $k_{399}$ is the wavenumber of the Bragg light.

A portion of the incoming atoms in the ground ${}^1{\mathrm{S}}_0$ state are selectively excited to the metastable ${}^3{\mathrm{P}}_2$ state via interaction with a continuous Gaussian excitation beam. The remaining atoms in the ground state are removed by a subsequently applied blast beam at \SI{399}{nm}. A de-excitation laser beam then selectively transfers the atoms in the metastable state back to the ground state in momentum space. The excitation and de-excitation laser beams exhibit Gaussian interaction profiles with identical beam waists \( w_e \). The transit time of \( w_e / v \) leads to a broadening of the excitation spectrum, where \( v \) denotes the atomic longitudinal velocity. The corresponding Gaussian root-mean-square (rms) momentum width is given by $\sigma_t = M v/(k_{507} w_e)$, where \( M \) is the atomic mass and \( k_{507} \) is the wavenumber of the 507~nm excitation and de-excitation lasers~\cite{demtroder_laser_1981}. The transverse momentum width of the final atomic beam can be expressed as $\sigma_t^{(\mathrm{out})} = \sigma_t/\sqrt{2+\big(\sigma_t/\sigma_t^{(\mathrm{in})}\big)^2}$, where \( \sigma_t^{(\mathrm{in})} \) denotes the rms transverse momentum width of the incident atomic beam.

\begin{figure}[h]
\includegraphics[scale=0.42]{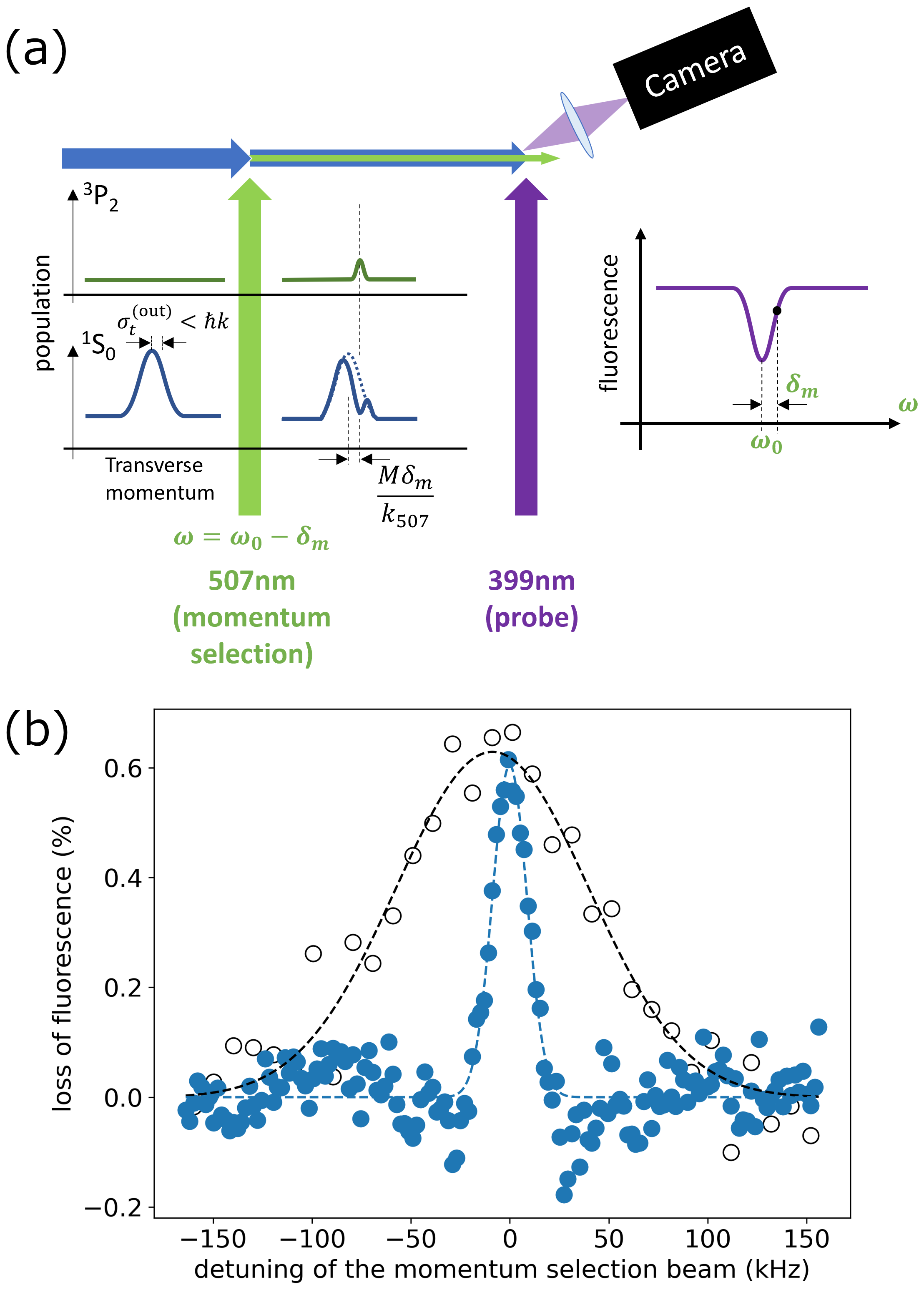}
\caption{\label{fig:fig2} (a) Schematic view of the momentum-resolving detection.
(b) Dependence of fluorescence loss on the detuning of the momentum-selection beam, shown without (black open circles) and with (blue filled circles) momentum filtering. Dashed lines indicate least-squares fits to the data.}
 \end{figure}

A technique for measuring the transverse momentum distribution of an atomic beam is essential not only for verifying momentum filtering, but also for the implementation of Bragg diffraction and atom interferometry.
Figure~\ref{fig:fig2}(a) illustrates the momentum-resolving detection scheme. A momentum selection beam, detuned by $\delta_m$ from the resonance frequency $\omega_0$ of the ultra-narrow \SI{507}{nm} transition, is applied orthogonally to the atomic beam.
The atomic beam with its transverse momentum of $M \delta_m / k_{507}$ is selectivley excited to the metastable state.
When a probe beam resonant with the dipole-allowed transition (\SI{399}{nm}) is subsequently applied to the atomic beam, fluorescence from the atoms decreases as a portion of the atoms transitions from the ground state to the metastable state. The overall momentum distribution can be extracted from the fluorescence reduction by scanning the frequency of the momentum selection beam.

For the demonstration of the method,
we prepared $^{171}{\rm Yb}$ atomic beam,
which requires less optical power for the transition
from ${}^{1}\mathrm{S}_0$ to ${}^{3}\mathrm{P}_2$~\cite{hosoya_high-flux_2023},
though essentially our method can be applied
to the bosonic isotopes, such as $^{174}{\rm Yb}$.
A continuous atomic beam of $^{171}{\rm Yb}$, characterized by a typical transverse momentum width of $4 \hbar k_{399}$ and a longitudinal velocity $v$ of approximately \SI{30}{m/s}, was generated through laser cooling using a combination of the dipole-allowed transition at \SI{399}{nm} and the ${}^{1}\mathrm{S}_0 - {}^{3}\mathrm{P}_1$ intercombination transition at \SI{556}{nm}, as described in~\cite{hosoya_high-flux_2023}. Note that the atomic beam comprised atoms in two nuclear spin sublevels ($I = 1/2$, $m_I = \pm 1/2$). In our setup, a magnetic field of $10^{-5}$~T is applied during the momentum filtering process, inducing Zeeman splitting of the magnetic sublevels. As a result, the filtering process selectively targets atoms in the $m_I = +1/2$ state, utilizing $\sigma_+$-polarized light for both excitation and de-excitation.
The beam waist of both the excitation and de-excitation beams was set to \( w_e = 1.1\,\mathrm{mm} \), which is expected to yield a filtered transverse momentum width of \( \sigma_t^{\mathrm{(out)}} = 0.27\,\hbar k_{399} \).
While it is preferable for the vertical beam waists of the excitation and de-excitation beams to exceed the $1/e^2$ radius (i.e., $2\sigma$) of the atomic beam, which is \SI{1.9}{mm} at the position of the laser beams, they were constrained to \SI{1.1}{mm}—the same as in the horizontal direction—due to the limited available laser power of \SI{500}{mW}.

To verify the effectiveness of our momentum filtering, we used a probe beam with a waist $w_p = \SI{0.3}{mm}$, much smaller than the atomic beam's $1/e^2$ radius of \SI{3.1}{mm} at the probe position.
In Fig.~\ref{fig:fig2}(b), the maximum fluorescence loss in the momentum-filtered case reaches that of the unfiltered one, indicating a nearly 100\% filtering efficiency. This high efficiency arises because the probe light interacted with atoms passing near the center of the excitation and de-excitation beams, where the $\pi$-pulse condition was satisfied. Figure~\ref{fig:fig2}(b) clearly shows that the filtering process effectively reduced the transverse momentum width of the atomic beam.
The Gaussian standard deviations of the measured spectra without and with filtering were $48.7(3.0)$~kHz and $8.7(0.4)$~kHz, respectively. After accounting for the transit-time broadening of 7.1~kHz associated with momentum-selective detection, the standard deviations of the spectra attributable to the atomic momentum width are estimated to be $48.2(3.0)$~kHz and $5.0(0.7)$~kHz, respectively. These correspond to transverse momentum widths of $4.3(0.3)\hbar k_{399}$ and $0.44(6)\hbar k_{399}$, respectively. The measured value exceeds the theoretical prediction of $0.27 \hbar k_{399}$,
calculated from the atomic beam velocity $v$ and the longitudinal beam waist $w_e$ of the excitation and de-excitation beams.
One possible cause of the discrepancy between the experimental results and the theoretical predictions is the presence of a residual AC magnetic field at the positions of the excitation and de-excitation beams. A field as small as $10^{-7}\,\mathrm{T}$ is sufficient to broaden the momentum width to $0.44\,\hbar k_{399}$.

To estimate the atomic flux after momentum filtering, we measured the fluorescence ratio with and without the filtering process, omitting the momentum selection step. A probe beam with a top-hat intensity profile was used, with vertical and horizontal dimensions of 7.5~mm and 2.5~mm, respectively. The vertical size exceeded the $1/e^2$ radius (3.1~mm) of the atomic beam, ensuring complete spatial overlap. The fluorescence ratio with and without filtering was measured to be $2.9(2)\%$. The total atomic flux was independently determined using conventional absorption spectroscopy to be $2.3(3) \times 10^8$~atoms/s. Based on the measured fluorescence ratio, the atomic flux after momentum filtering was estimated to be $6.7(9) \times 10^6$~atoms/s. Assuming ideal filtering within a transverse momentum window of $0.44\,\hbar k_{399}$ from an initial momentum distribution width of $4.3\,\hbar k_{399}$, the expected atomic flux would be $1.2(2) \times 10^7$~atoms/s. The experimentally observed filtering efficiency of $56(5)\%$ is expected to approach unity with increased optical power and larger vertical beam waists for both excitation and de-excitation beams.

\begin{figure}[h]
    \includegraphics[scale=0.45]{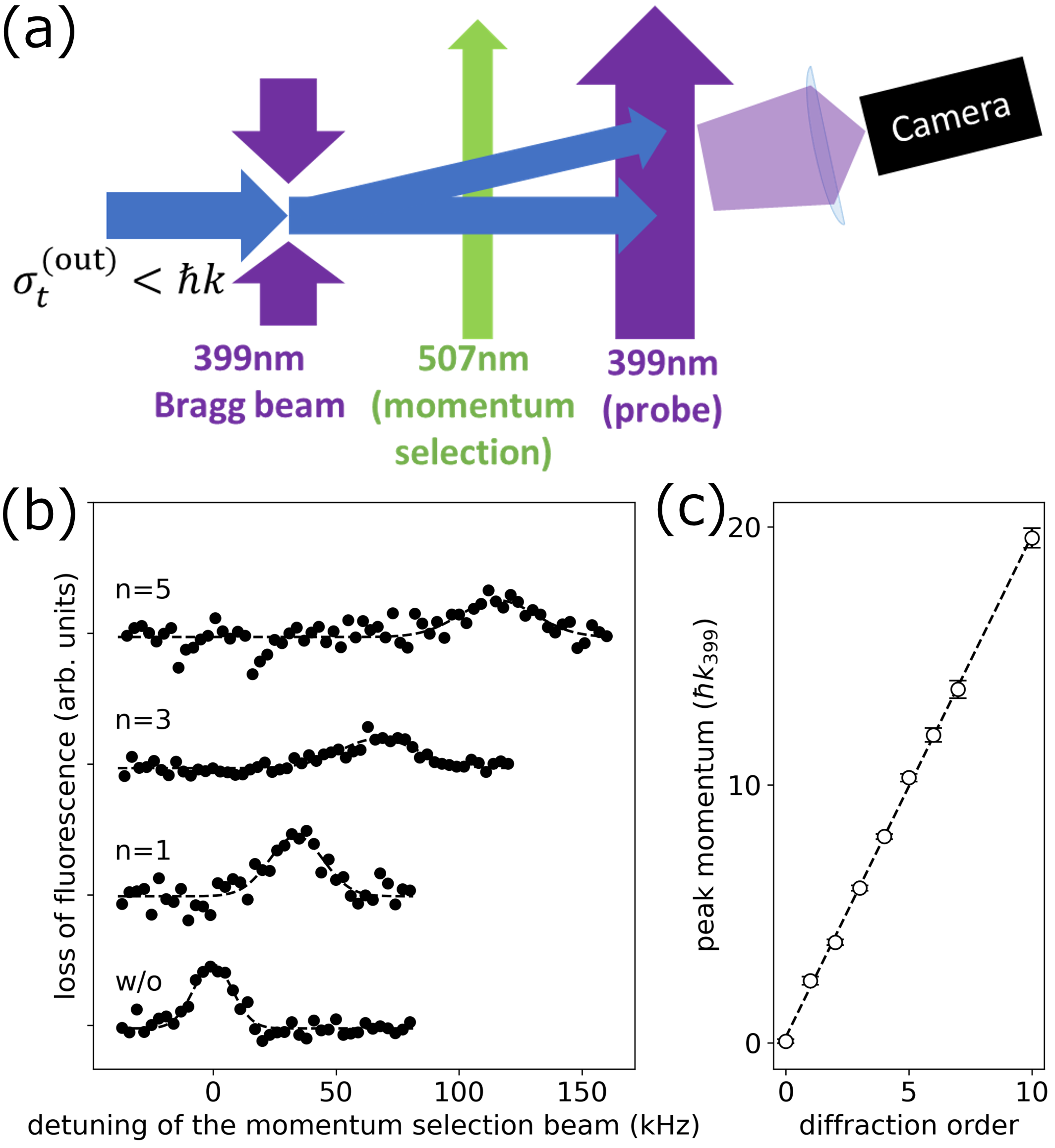}
    \caption{\label{fig:fig3} (a) Schematic view of the setup for the Bragg diffraction experiment. (b) Obtained loss in the momentum-resolving detection for each setting for the diffraction order. Dashed lines are fitting curves obtained by the least-square method. (c) The dependence of the momentum transfer on the Bragg order. }
\end{figure}

We observed quasi-Bragg diffraction using a cold Yb atomic beam with a transverse momentum width below the recoil limit.
Figure~\ref{fig:fig3}(a) shows the experimental setup for the Bragg diffraction.
Bragg beams, consisting of a pair of counter-propagating laser beams with a frequency difference of $\delta_n$, are applied orthogonally to the incoming cold atomic beam, where $\delta_n=4 n \omega_r$,
with $n$ denoting the diffraction order and $\omega_r$ the recoil frequency.
Quasi-Bragg diffraction imparts a momentum transfer of 
$2n\hbar k_{399}$ to the atom, while leaving the atom in its internal ground state.
The post-diffraction momentum distribution was measured via the momentum-selective detection scheme illustrated by Fig.~\ref{fig:fig2}(a).
The horizontal and vertical waists of the Bragg beam were \SI{0,25}{mm} and \SI{2.5}{mm}, respectively.
The detuning was set to \SI{-20}{GHz} relative to the ${}^1\mathrm{S}_0$–${}^1\mathrm{P}_1$ transition.
The laser power was optimized to maximize the efficiency of each Bragg diffraction order; for example, a power of \SI{25}{mW} was used for third-order quasi-Bragg diffraction.
The narrow transverse momentum width of the atomic beam allows individual diffraction orders to be resolved (Figs.\ref{fig:fig3}~(b) and (c)).

\begin{figure}[h]
\includegraphics[scale=0.50]{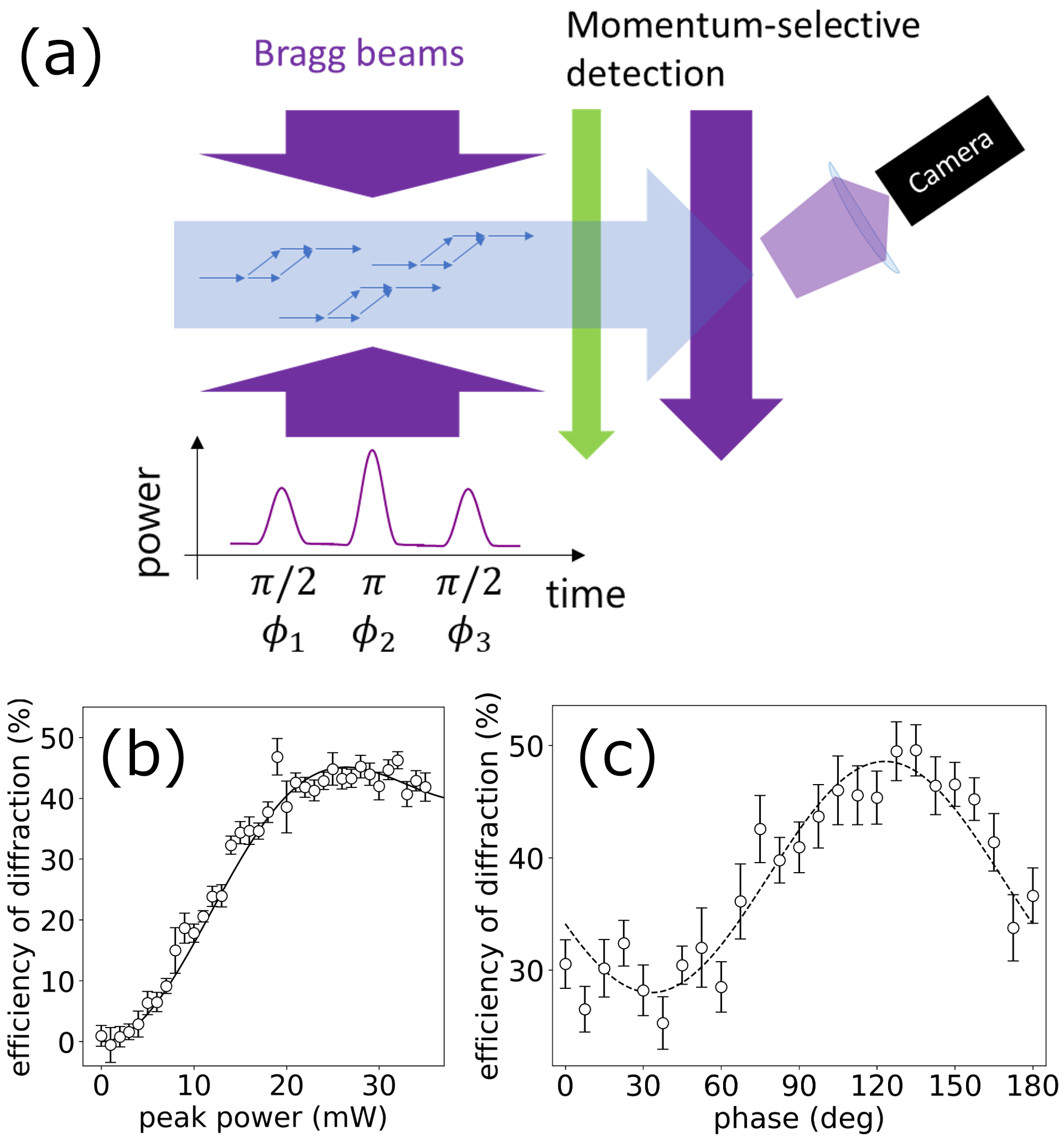}
\caption{\label{fig:fig4} (a) Schematic view of the setup for the Bragg interferometer experiment.
(b) The dependence of the diffraction efficiency on the power of Bragg light. The solid line is the theoretical calculation for current experimental parameteres.
(c)  The dependence of the population in $2 \hbar k_{399}$ momentum state
    on the phase of the central $\pi$ pulse.
    Dashed line is the least-square fitting with a sinusoidal function.
}
\end{figure}

Finally, we demonstrated a Mach–Zehnder interferometer (MZI) based on quasi-Bragg diffraction using a momentum-filtered atomic beam (see Fig.~\ref{fig:fig4}(a)). While spatial-domain MZIs with continuous atomic beams require three phase-synchronized Bragg beams, a time-domain MZI~\cite{torii_mach-zehnder_2000} can be realized with a single beam. We adopted this simpler approach for our demonstration with a slow atomic beam.
The Bragg light field was formed by counter-propagating top-hat beams with a frequency offset. A precisely timed sequence of \(\pi/2\text{--}\pi\text{--}\pi/2\) pulses acted as the beam splitter and mirror, completing the interferometer. Since uniform wavefronts and intensity over the entire sequence are essential to ensure coherent manipulation of the atomic ensemble, we used a beam shaper to generate a 7.5~mm-diameter top-hat beam, similar to the method in~\cite{mielec_atom_2018}.
The $\pi/2$--$\pi$--$\pi/2$ pulse sequence had to be completed within a duration much shorter than the time required for the atoms to traverse the entire Bragg beam area (\SI{250}{\micro\second}). To satisfy this condition, a pulse with a standard deviation of \SI{3.5}{\micro\second} was employed, although the optimal pulse width for quasi-Bragg diffraction is estimated to be \SI{7}{\micro\second}. The time interval between pulses was set to \SI{24.5}{\micro\second}, resulting in a total sequence duration of \SI{73.5}{\micro\second}. To increase the number of atoms participating in the MZI, the $\pi/2$-$\pi$-$\pi/2$ pulse sequence was repeated every \SI{250}{\micro\second}.
The temporal profile of the optical pulses was controlled using acousto-optic modulators (AOMs) placed in both counter-propagating Bragg beams.

We first observed Rabi oscillations between $0\hbar k_{399}$ and $2\hbar k_{399}$ using a single Bragg pulse (Fig.~\ref{fig:fig4}(b)), along with momentum-resolved detection as described previously. The observed power dependence of the population was consistent with theoretical calculations, allowing us to estimate the appropriate laser powers for the $\pi/2$ and $\pi$ pulses. In the experiment, we employed powers of \SI{15}{mW} and \SI{20}{mW} with a detuning of $-4.5$~GHz, while numerical calculations predicted optimal powers of \SI{15}{mW} and \SI{22}{mW}, respectively. The output phase of the MZI, \(\phi\), consists of the inertial phase shift \(\phi_{\mathrm{inertial}}\), arising from acceleration and rotation, and the phases of the three Bragg pulses, \(\phi_1\), \(\phi_2\), and \(\phi_3\). These are related by
$\phi = \phi_{\mathrm{inertial}} + \phi_1 - 2\phi_2 + \phi_3.$
When the apparatus is stationary in the laboratory frame, \(\phi_{\mathrm{inertial}}\) remains constant. By varying the phase of the central \(\pi\) pulse, \(\phi_2\), from \(0^\circ\) to \(180^\circ\)—achieved by adjusting the relative phase between the two RF signals driving the AOMs—we observed a corresponding modulation in the momentum populations (see Fig.~\ref{fig:fig4}(c)).
The interference fringe, observed via oscillations between $0 \hbar k_{399}$ and $2 \hbar k_{399}$, varied at twice the phase of the second pulse, in agreement with theoretical predictions.
The measured 20(2)\% interference contrast is consistent with the theoretical prediction of 23(3)\%, estimated based on the single-pulse transition efficiency and the repetition duty cycle of the pulse sequence.

In conclusion, we implemented momentum filtering using an ultra-narrow optical transition to generate an atomic beam with a sub-recoil transverse momentum width, suitable for Bragg interferometry. Applied to a laser-cooled, slow, and continuous Yb atomic beam (\SI{30}{m/s}, flux of $2.3(3) \times 10^8$~atoms/s), the filtering yielded a width of $0.44(6) \hbar k_{399}$ and a flux of $6.7(9) \times 10^6$~atoms/s. Using a momentum-resolved detection scheme, we achieved higher order quasi-Bragg diffraction and demonstrated a time-domain Mach–Zehnder interferometer employing a momentum-filtered atomic beam. These results support the use of continuous, sub-recoil atom sources for precision interferometry and pave the way for inertial sensors based on alkaline-earth(-like) atoms that are inherently insensitive to magnetic field fluctuations.

\section*{Acknowledgments}
This work was supported by JST, Japan Grant Numbers JPMJMI17A3 and JPMJPF2015.
T.H. acknowledges partial support from the Japan Society for the Promotion of Science.

\bibliographystyle{apsrev4-2}
\bibliography{Cold_atomic_beam_Bragg_interferometer_cleaned}

\end{document}